\documentclass[11pt]{article}

\usepackage[T1]{fontenc}
\usepackage[utf8]{inputenc}
\usepackage[margin=1in]{geometry}
\usepackage{lmodern}
\usepackage{microtype}
\usepackage{amsmath,amssymb}
\usepackage{graphicx}
\usepackage{booktabs}
\usepackage{placeins}
\usepackage[numbers,sort&compress]{natbib}
\usepackage[colorlinks=true,citecolor=blue,linkcolor=blue,urlcolor=blue]{hyperref}

\newcommand{\Parasplit}{}
\newcommand{\numberthis}{\addtocounter{equation}{1}\tag{\theequation}}
\newcommand{\JournalTitle}[1]{#1}
\newcommand{\matmethods}[1]{\section*{Materials and Methods}#1}
\newcommand{\showmatmethods}[1]{}
\newcommand{\dataavail}[1]{\section*{Data Availability}#1}
\newcommand{\acknow}[1]{\section*{Acknowledgments}#1}
\newcommand{\showacknow}[1]{}

\title{\textbf{Widespread plasma jets in the Martian magnetosheath revealed by dual-spacecraft observations}}

\author{%
Abigail Tadlock$^{1}$, Chuanfei Dong$^{1,2,*}$, Chi Zhang$^{1,*}$, Savvas Raptis$^{3}$,\\
Hongyang Zhou$^{1}$, Xinmin Li$^{1}$, and Jiawei Gao$^{1}$\\[0.75em]
\small $^{1}$Department of Astronomy, Center for Space Physics, Boston University, Boston, MA 02215\\
\small $^{2}$School of Natural Sciences, Institute for Advanced Study, Princeton, NJ 08540\\
\small $^{3}$The Johns Hopkins University Applied Physics Laboratory, Laurel, MD 20723
}
\date{}

\begin{document}
\maketitle
\begingroup
\renewcommand{\thefootnote}{\fnsymbol{footnote}}
\footnotetext[1]{\href{mailto:dcfy@bu.edu}{dcfy@bu.edu}; \href{mailto:zc199508@bu.edu}{zc199508@bu.edu}.}
\endgroup

\begin{abstract}
Plasma jets in planetary magnetosheaths are transient, high dynamic pressure structures capable of strongly perturbing downstream magnetospheres and modulating solar wind energy transfer. Although recently identified at Mars, their formation mechanisms and global occurrence remain poorly constrained. Using dual-spacecraft observations from Mars Atmosphere and Volatile EvolutioN (MAVEN) and Tianwen-1, we show that magnetosheath jets occur throughout the Martian magnetosheath, in striking contrast to Earth where jet occurrence is strongly controlled by interplanetary magnetic field orientation and largely confined to quasi-parallel shock regions. We propose that this widespread distribution is linked to Mars' extended hydrogen corona, which generates pickup ions that drive proton-cyclotron waves upstream of the bow shock. Because this mechanism is generic to any body with an extended corona (or exosphere) embedded in a stellar wind, this finding bears on solar-wind interactions at comets and other small bodies, and on the space environments of unmagnetized terrestrial exoplanets. These results indicate that magnetosheath jet formation can be governed not only by external solar wind forcing but also by intrinsic planetary properties, pointing to a previously unrecognized pathway for solar wind-planet coupling at unmagnetized worlds.
\end{abstract}

At Earth, transient plasma jets behind the bow shock are a key mediator of solar wind--magnetosphere coupling, capable of driving magnetopause deformation \cite{ma_comprehensive_2024,amata_high_2011} and surface waves \cite{archer_direct_2019}, induced ground currents \cite{archer_magnetospheric_2013}, and dayside aurora \cite{wang_impacts_2018}. These so-called magnetosheath jets are brief, localized enhancements in dynamic pressure that can exceed the ambient solar wind \cite{plaschke_jets_2018,kramer_jets_2024}, and geoeffective ones occur several times per hour \cite{plaschke_geoeffective_2016}, making it important to understand how and where they form.

Magnetosheath jets originate at planetary bow shocks \cite{hietala_supermagnetosonic_2009}. At Earth, extensive multipoint observations show that jets occur predominantly downstream of the quasi-parallel (Qpar) bow shock \cite{plaschke_anti-sunward_2013,vuorinen_jets_2019,raptis_classifying_2020}, where the angle $\theta_{Bn}$ between the bow shock normal and the interplanetary magnetic field (IMF) is small ($\theta_{Bn}<45^\circ$). Jet formation has been linked to upstream structures such as foreshock transients and IMF discontinuities \cite{archer_magnetosheath_2012,suni_magnetosheath_2025}, as well as to nonstationary bow shock processes associated with steepened foreshock waves, including Short Large Amplitude Magnetic Structures (SLAMS) \cite{hietala_generation_2013, raptis_downstream_2022}. Whether this IMF-controlled paradigm applies universally across planetary systems remains largely unexplored. Although magnetosheath jets have recently been reported at Mars, their formation and global occurrence remain poorly understood \cite{gunell_magnetosheath_2023,zhou_magnetosheath_2024}.

Mars provides a natural laboratory for exploring jet formation at weakly magnetized or unmagnetized planets. Unlike Earth, Mars lacks a global intrinsic magnetic field and instead hosts an induced magnetosphere shaped by solar wind and IMF draping around the ionosphere \cite{zhang_three-dimensional_2022,gao_two_2024} and localized crustal magnetic fields \cite{nagy_plasma_2004,dong_solar_2015}. Initial observations suggest that Martian jets differ from terrestrial jets: they are primarily density-driven, with large density enhancements and relatively modest velocity \Parasplit
 increases \cite{mohammed-amin_jets_2025}, whereas terrestrial jets often exhibit strong velocity enhancements \cite{raptis_classifying_2020}. These differences point to formation mechanisms that may not have a direct Earth analog -- most plausibly determined by Mars' planetary properties and the features of its near-space environment.

Mars possesses an extended hydrogen corona (or exosphere) that extends far into the solar wind, up to $\sim10$ $R_M$ (Mars radii) \cite{nagy_plasma_2004,wei_proton_2006}. Ionization of these hydrogen atoms produces pickup ions that generate proton-cyclotron waves (PCWs) upstream of the bow shock \cite{zhang_role_2025,mazelle_bow_2004,delva_upstream_2011,romanelli_proton_2016}. Unlike reflected ion foreshock waves, this wave activity can occur across a broad range of $\theta_{Bn}$ rather than being confined to upstream of the Qpar shock, modifying shock dynamics and potentially enabling jet formation over a wider range of $\theta_{Bn}$ than at Earth.

\begin{figure}
    \centering
    \includegraphics[width=1\linewidth]{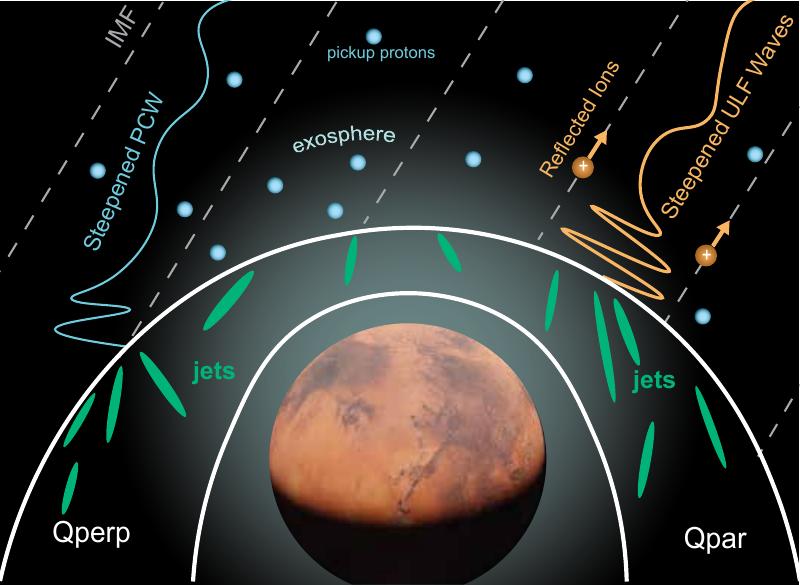}
    \caption{\textbf{Schematic illustration of magnetosheath jet formation at Mars.}
    The Qpar shock is shown on the right and the Qperp shock on the left. Upstream of the Qpar shock, reflected ions backstream along the interplanetary magnetic field (IMF), generating ultra--low-frequency (ULF) waves (orange curve) that steepen toward the shock (white curve). Upstream of the Qperp shock, pickup protons (light blue), originating from the extended Martian hydrogen corona, generate PCWs, which likewise steepen (light blue curve). Interaction of these steepened waves with the bow shock can lead to magnetosheath jet formation. Additional processes, including nonstationary shock dynamics, may also contribute and help explain the widespread occurrence of jets across the Martian magnetosheath.}
    \label{fig:sk}
\end{figure}

Simultaneous two-point observations at Mars from MAVEN \cite{jakosky_mars_2015} in the magnetosheath and Tianwen-1 \cite{zou_scientific_2021} in the upstream solar wind now enable direct investigation of the solar wind–jet connection. Using these observations, we present the first global assessment of magnetosheath jet occurrence at Mars. We show that, in contrast to Earth, jet occurrence spans the full range of $\theta_{Bn}$, with enhanced occurrence at $\theta_{Bn}>45^\circ$ near southern summer solstice, when exospheric PCW activity peaks \cite{romeo_variability_2021}. We attribute this enhanced jet activity to the interaction of steepened PCWs with the quasi-perpendicular (Qperp) shock, suggesting a jet formation mechanism distinct from that at Earth (Fig.~\ref{fig:sk}). Beyond Mars, this mechanism -- exosphere-driven wave activity modifying shock dynamics -- is generic to any unmagnetized body with an extended exosphere in a stellar wind, with implications for the space environments of comets and terrestrial exoplanets.

\section*{Results}
\subsection*{Representative Jets}

\begin{figure*}[t]
    \centering
    \includegraphics[width=0.95\textwidth]{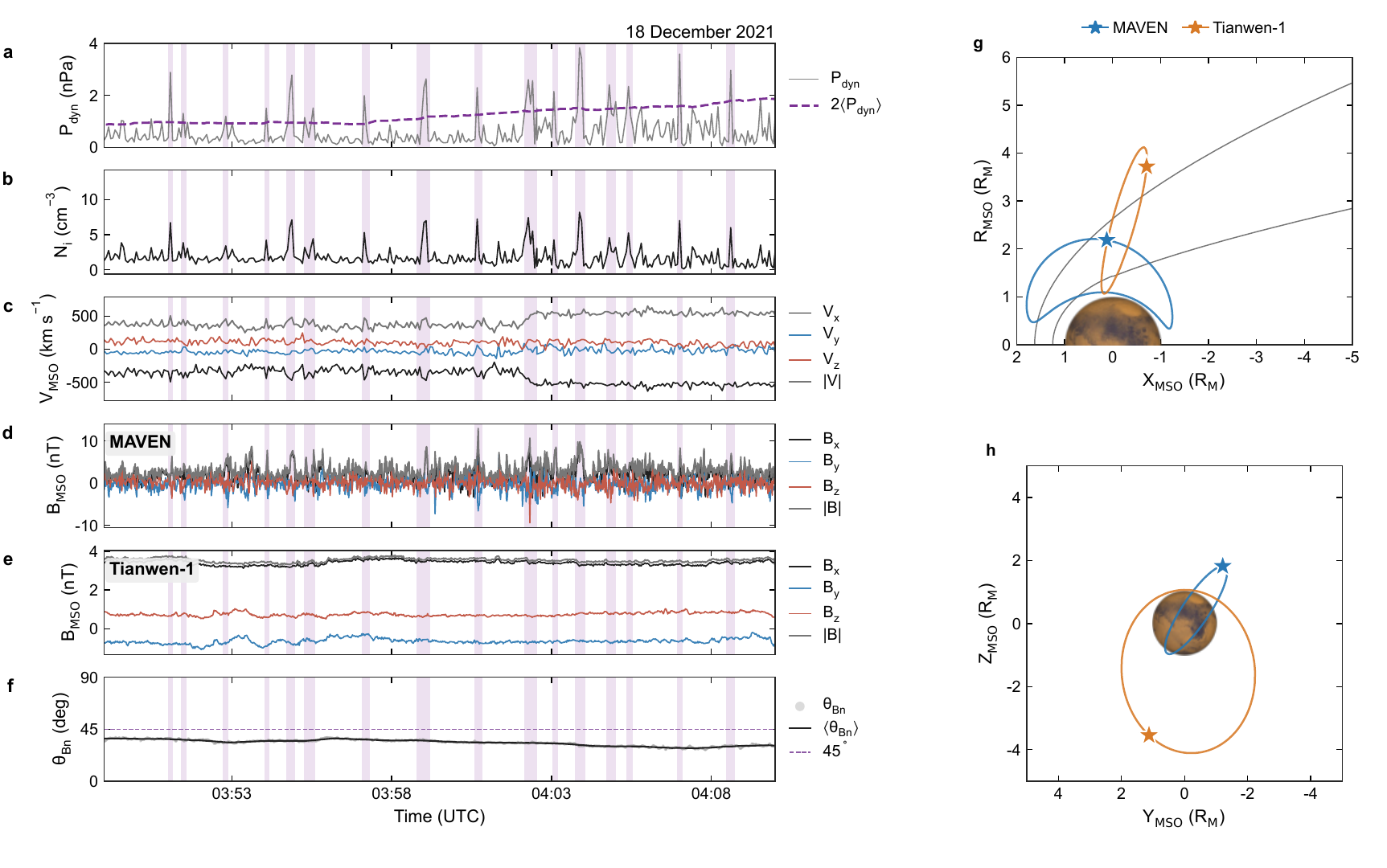}
    \caption{\textbf{Magnetosheath jets observed in the Martian Qpar magnetosheath.}
    (a) Dynamic pressure. The purple dashed curve represents twice the local background dynamic pressure,
    $2\left\langle P_{\mathrm{dyn}} \right\rangle$, where
    $\left\langle P_{\mathrm{dyn}} \right\rangle$ is calculated using a centered 10-min moving average.
    (b) Ion density measured by the Solar Wind Ion Analyzer (SWIA).
    (c) Ion velocity measured by SWIA.
    (d) Magnetic field measured by MAVEN.
    (e) Magnetic field measured by Tianwen-1.
    (f) $\theta_{Bn}$ calculated using the IMF measured by Tianwen-1 and the position of MAVEN; the black curve shows a 1-min moving average.
    (g) Trajectories and positions of MAVEN (blue) and Tianwen-1 (orange) in the cylindrical X–R plane.  The stars mark the average spacecraft positions during the displayed interval.
    (h) Positions of MAVEN and Tianwen-1 in the Y–Z MSO plane.
    }
    \label{fig:par}
\end{figure*}

\begin{figure*}[t!]
    \centering
    \includegraphics[width=0.95\textwidth]{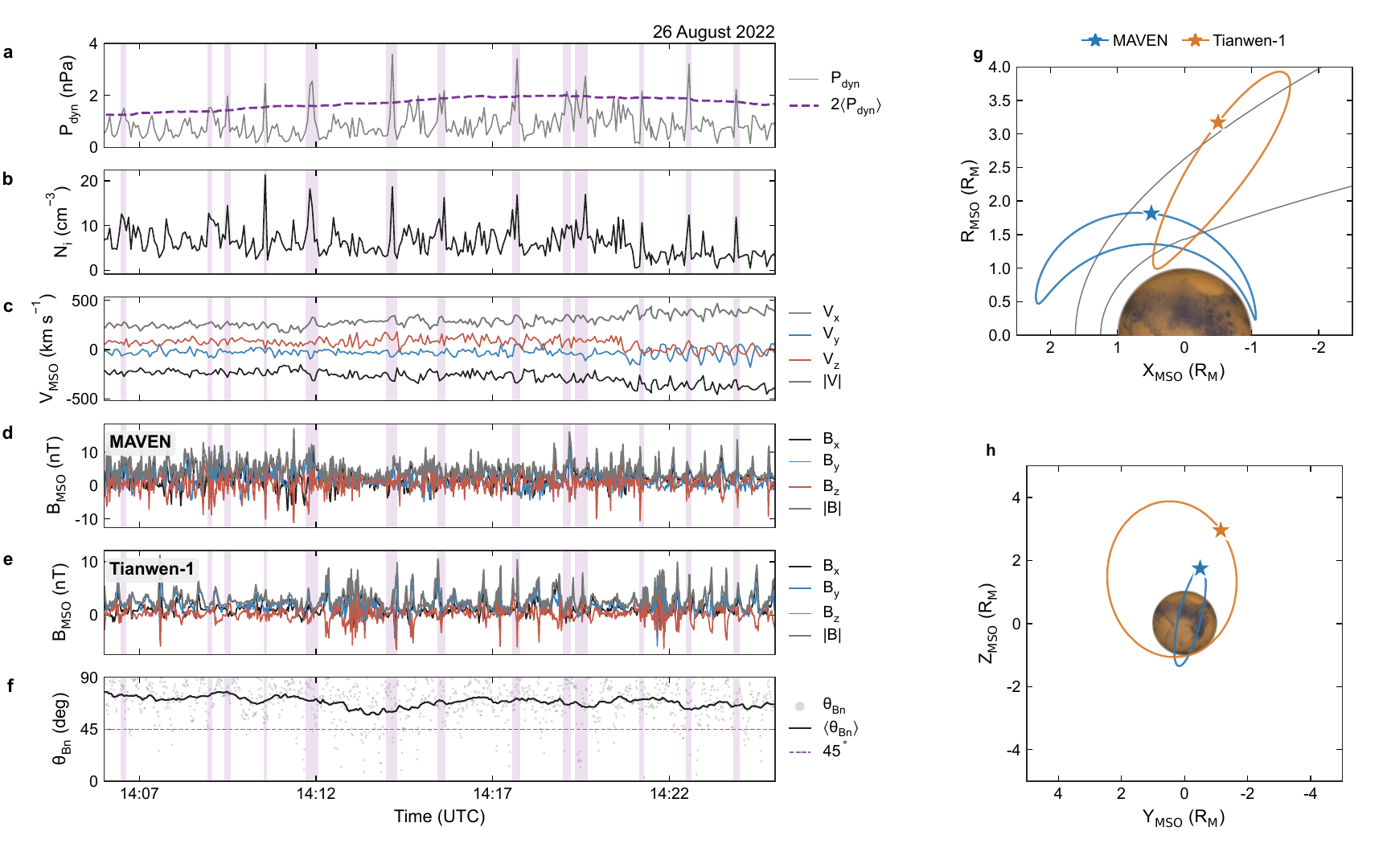}
    \caption{\textbf{Magnetosheath jets observed in the Martian Qperp magnetosheath.}
	Same format as Fig.~\ref{fig:par}, showing a similar event on 26 August 2022. See Fig.~\ref{fig:par} for panel definitions.
    }
    \label{fig:perp}
\end{figure*}

We highlight two events in which a series of magnetosheath jets were observed over tens of minutes, occurring in both Qpar ($\theta_{Bn}<45^\circ$, Fig. \ref{fig:par}) and Qperp ($\theta_{Bn}>45^\circ$, Fig. \ref{fig:perp}) magnetosheaths. These events are notable for the large number of jets and their possible periodic occurrence. Despite different $\theta_{Bn}$, both events share similar characteristics: jets are predominantly density-driven, with several-fold enhancements above the background, show modest velocity increases, and are co-located with magnetic field enhancements.

Fig. \ref{fig:par} shows jets in the Qpar magnetosheath on 18 December 2021 between 03:49 and 04:10 UTC. During this event, MAVEN was moving outward from Mars toward the bow shock (Fig.~S2). It observed a series of jets with an approximately one-minute interval between peaks (Fig. \ref{fig:par}a). Most jets coincided with magnetic field enhancements (Fig. \ref{fig:par}d) and modest velocity increases (Fig. \ref{fig:par}c). The jets were primarily driven by strong density enhancements (Fig. \ref{fig:par}b), several times the background level. Tianwen-1 was located in the opposite Z-hemisphere from MAVEN (Fig.~\ref{fig:par}h), outside the foreshock region, where it observed a steady IMF, consistent with the formation of a foreshock upstream of the Qpar shock. These jets may therefore be associated with foreshock SLAMS or other compressive magnetic structures, discussed further in the next section.

Fig. \ref{fig:perp} shows jets in the Qperp magnetosheath on 26 August 2022 between 14:06 and 14:25 UTC. MAVEN observed multiple density-driven jets accompanied by modest velocity increases and pronounced magnetic field enhancements, particularly after 14:20 UTC. Tianwen-1 was located upstream of the shock in the same region, approximately 1.6 $R_M$ away, and observed similar compressive magnetic structures consistent with SLAMS (Fig.~\ref{fig:perp}e). Most magnetic field enhancements observed by MAVEN corresponded to some increase in dynamic pressure, although not all were strong enough to be classified as jets (only 3 of the 8 SLAMS observed after 14:20 were associated with jets). Most jets corresponded to peaks in magnetic field strength observed by Tianwen-1 (Fig.~\ref{fig:perp}e), particularly the more distinct SLAMS after 14:20 UTC.

We suggest that these jets are SLAMS transmitted through the bow shock. During this interval $\theta_{Bn}$ was approximately $70^\circ$, indicating Qperp conditions (Fig. \ref{fig:perp}f).  At Earth, SLAMS form predominantly in the foreshock upstream of the Qpar shock. At Mars, however, SLAMS have been observed at high $\theta_{Bn}$, where they can arise from PCWs generated by pickup ions in the extended hydrogen exosphere \cite{zhang_role_2025}. This event occurred at a solar longitude ($L_s$) of $292^\circ$, near Martian perihelion ($251^\circ$) and the peak of exospheric column density ($\sim270^\circ$) \cite{halekas_seasonal_2017,clarke_martian_2024}, when pickup-ion-driven wave activity is expected to be enhanced.

\begin{figure}[t!]
    \centering
    \includegraphics[height=0.72\textheight,keepaspectratio]{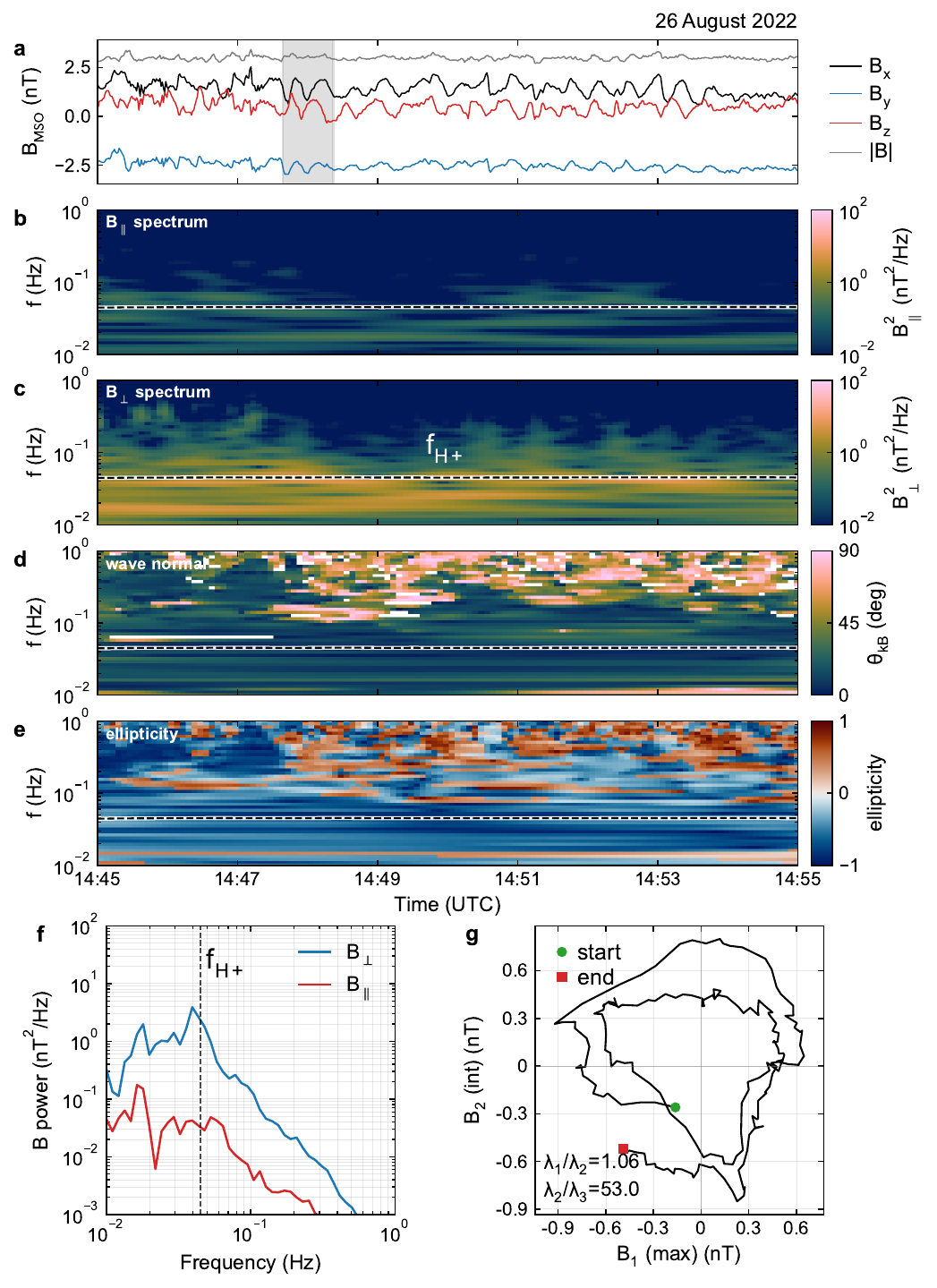}
    \caption{\textbf{PCWs observed by MAVEN.}
    (a) Magnetic field components ($B_x$, $B_y$, $B_z$, and $|\mathbf{B}|$).
    The gray-shaded sub-interval marks the window used for the minimum
    variance analysis (MVA) in panel (g).
    (b) $B_\parallel$ wave power spectrum.
    (c) $B_\perp$ wave power spectrum.
    (d) Wave normal angle, defined as the angle between the wave
    propagation direction and the background magnetic field.
    (e) Wave ellipticity (negative values indicate left-hand polarization
    in the spacecraft frame).
    The black dashed line in panels (b)--(e) marks the proton
    gyrofrequency, $f_{\rm H^+}$.
    (f) Median power spectral density of the transverse ($B_\perp$, blue)
    and compressive ($B_\parallel$, red) magnetic field components over
    the 14:45--14:55 UTC window; the vertical dashed line indicates
    $f_{\rm H^+}$.
    (g) Hodogram of the magnetic field in the maximum--intermediate
    variance plane from MVA over the gray-shaded sub-interval in panel
    (a) (14:47:39 -- 14:48:22 UTC, $\sim 2$ proton-cyclotron
    periods). The eigenvalue ratios $\lambda_1/\lambda_2$ and $\lambda_2/\lambda_3$ are given in the
    bottom-left corner. The trace runs from the green dot (start) to
    the red square (end).
    }
    \label{fig:pcw}
\end{figure}

MAVEN crossed the bow shock shortly after the interval shown in Fig.~\ref{fig:perp}, entering the upstream solar wind. There, it observed waves consistent with PCWs (Fig.~\ref{fig:pcw}). These waves exhibit power concentrated primarily in the transverse ($B_\perp$) component (Fig.~\ref{fig:pcw}c) with a peak frequency close to the local proton gyrofrequency $f_{\rm H^+}$ (Fig.~\ref{fig:pcw}f), low wave-normal angle indicating propagation along the background magnetic field (Fig.~\ref{fig:pcw}d), and negative ellipticity corresponding to left-hand polarization in the spacecraft frame (Fig.~\ref{fig:pcw}e). Minimum variance analysis (MVA) over the gray-shaded sub-interval in Fig.~\ref{fig:pcw}a ($\sim 2$ proton-cyclotron periods within the wave packet) yields a hodogram in the maximum--intermediate variance plane (Fig.~\ref{fig:pcw}g) with eigenvalue ratio $\lambda_1/\lambda_2 = 1.06$, demonstrating near-circular polarization. $\lambda_2/\lambda_3=53.0$, indicating that the MVA clearly distinguished between the minimum and intermediate variance directions. With the MVA basis aligned so that the wave-normal direction $\hat{e}_3$ points along the background magnetic field, $\mathbf{B}_0$, the trace rotates clockwise from start (green dot) to end (red square), which corresponds to left-hand polarization in the spacecraft frame and is consistent with the negative ellipticity at $f_{\rm H^+}$ in Fig.~\ref{fig:pcw}e. The combination of near-circular, left-hand polarization, quasi-parallel propagation, and a transverse-dominated spectrum peaked just below $f_{\rm H^+}$ matches the canonical PCW signature \cite{zhang_role_2025, romanelli_proton_2016}. For the duration of this event, $\theta_{Bn} > 70^\circ$ (see Fig.~S3), further supporting the PCW interpretation, as foreshock ULF waves are generally not expected upstream of the Qperp shock. The presence of upstream PCWs supports the interpretation that the jets observed in Fig.~\ref{fig:perp} originate from steepened PCWs. Longer intervals showing MAVEN bow shock crossings for both events are provided in Figs.~S2 and S3.

\subsection*{Dependence on solar wind parameters and $\theta_{Bn}$}\label{ssec:sw}
\begin{figure*}
    \centering
    \includegraphics[width = 0.9\linewidth]{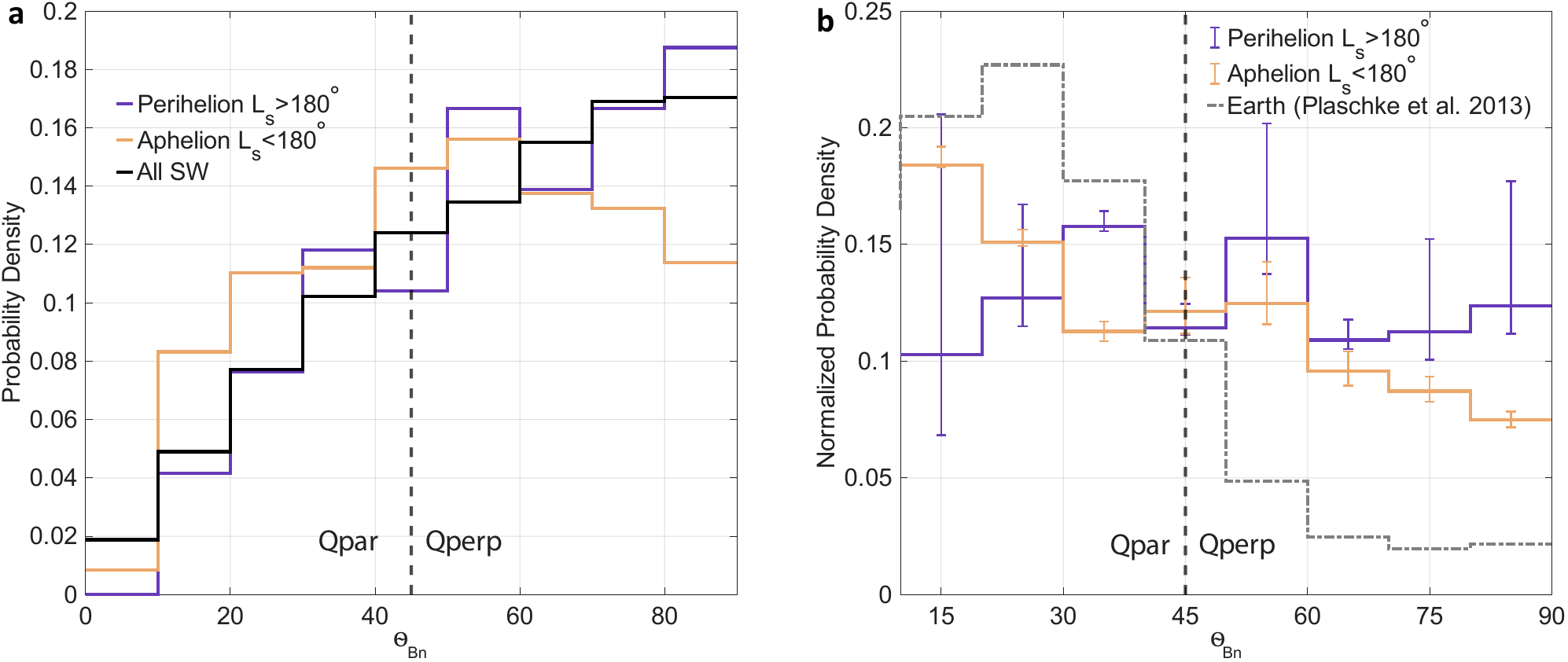}
    \caption{\textbf{Jet occurrence as a function of $\theta_{Bn}$.}
    (a) Probability density of jet occurrence at Mars versus $\theta_{Bn}$. The dataset is separated into near-perihelion ($L_s>180^\circ$, when exospheric density and PCW occurrence peak) and near-aphelion ($L_s<180^\circ$). The black line shows the distribution of all magnetosheath observations, which reflects MAVEN's sampling of the Qperp and Qpar regions.
    (b) Normalized probability density of jet occurrence versus $\theta_{Bn}$, shown for the same near-perihelion and near-aphelion subsets and corrected for MAVEN's sampling bias. Data below $10^\circ$ are excluded due to limited coverage. Bootstrapped $3\sigma$ confidence intervals are shown. The gray dotted line shows the corresponding distribution at Earth from \cite{plaschke_anti-sunward_2013}.
    }
    \label{fig:theta}
\end{figure*}

To further explore the factors controlling jet formation at Mars, we performed a statistical study of jets as a function of upstream solar wind parameters. Jets occur preferentially under intense solar wind conditions (Fig. S1). In particular, jet occurrence increases with higher solar wind dynamic pressure and velocity, lower IMF strength, and higher Alfv\'en Mach number, while showing little dependence on solar wind density. Jets also occur preferentially at lower IMF cone angles, which appears to favor jet formation, especially for Qpar jets. It is interesting to note that PCW occurrence also peaks at low to intermediate IMF cone angles \cite{romeo_variability_2021}. These trends are broadly similar for Qpar and Qperp jets. 

Fig. \ref{fig:theta}a shows the probability density histogram of jet occurrence as a function of $\theta_{Bn}$, separated by events closer to perihelion ($L_s > 180^\circ$) and aphelion ($L_s < 180^\circ$). This serves as a first-order approximation for PCW occurrence, which peaks near $L_s\approx270^\circ$ (the southern summer solstice, slightly later than perihelion at $L_s=251^\circ$) and has elevated occurrence at $L_s>180^\circ$ \cite{romeo_variability_2021}. The ``All SW'' dataset shown in Fig. \ref{fig:theta}a was assembled by identifying all periods when MAVEN was located in the nominal magnetosheath and Tianwen-1 in the solar wind, as determined by the model \cite{trotignon_martian_2006}. MAVEN observes the Qperp magnetosheath more frequently because the nominal Parker spiral angle at Mars is approximately $57^\circ$ \cite{liu_statistical_2021}. To account for this observational bias, the jet probability density function (PDF) was divided by the PDF of the ``All SW'' dataset in Fig.~\ref{fig:theta}b. The resulting distribution was then renormalized so that its integral equals unity. The ``All SW'' dataset was separated by solar longitude in the same manner as the jet dataset (see Fig.~S4). Fig. \ref{fig:theta}b shows the corresponding normalized distributions; data below $10^\circ$ are omitted due to poor data coverage. The occurrence of Martian jets is relatively uniform across $\theta_{Bn}$. For $L_s < 180^\circ$, jets are most common at low $\theta_{Bn}$, with their occurrence decreasing as $\theta_{Bn}$ increases, similar to Earth (see the gray dotted line in Fig. \ref{fig:theta}b), indicating that IMF orientation somewhat controls jet formation. However, the trend is much weaker than at Earth, implying additional factors contribute to jet formation in the Qperp magnetosheath at Mars. For $L_s > 180^\circ$, occurrence is relatively flat with little variation with $\theta_{Bn}$. Jets become more common in the high $\theta_{Bn}$ region when closer to perihelion. This is in contrast to Earth, where the vast majority of jets occur at $\theta_{Bn}<45^\circ$. These results suggest that, in addition to solar wind conditions, other processes influence jet formation, particularly near perihelion, which will be discussed further in the following section.

\section*{Discussion}
Magnetosheath jets at Mars are widespread and show only a weak dependence on $\theta_{Bn}$. This is notably different from Earth, where jets occur predominantly at low $\theta_{Bn}$. This difference suggests that jet formation at weakly magnetized Mars is influenced by processes absent at Earth, most likely related to the extended Martian hydrogen corona (or exosphere). Jet occurrence shows a seasonal variation, with Qperp jets becoming more frequent near perihelion, when the exosphere is more extended \cite{yamauchi_seasonal_2015, clarke_martian_2024,halekas_seasonal_2017} and PCW activity is enhanced \cite{romanelli_proton_2016,romeo_variability_2021}. We therefore propose that upstream wave activity, which can steepen into SLAMS, likely plays an important role in generating jets across the bow shock at Mars.

SLAMS have long been proposed as a potential formation mechanism of magnetosheath jets at Earth, and recent studies have linked them to jets \cite{raptis_downstream_2022,karlsson_origin_2015}. The events shown in Figs.~\ref{fig:par} and \ref{fig:perp} are consistent with jets driven by SLAMS. In the Qperp event (Fig.~\ref{fig:perp}), Tianwen-1 observed SLAMS upstream of the bow shock, while MAVEN detected similar structures co-located with jets in the magnetosheath, suggesting that these structures were transmitted through the shock. Furthermore, MAVEN observed PCWs upstream of the Qperp shock (Fig. \ref{fig:pcw}), supporting the idea that the SLAMS originated from PCWs. For the Qpar event (Fig.~\ref{fig:par}), Tianwen-1 was not located in the foreshock or near MAVEN and thus did not observe associated upstream structures, limiting direct identification of their source.

Comparing the solar wind drivers of jet formation at Mars and Earth shows both similarities and important differences. At Mars, jets occur preferentially during periods of high solar wind dynamic pressure (Fig. S1), consistent with terrestrial observations \cite{lamoury_solar_2021, vuorinen_solar_2023}. Jets at Earth exhibit a strong dependence on $\theta_{Bn}$, occurring five to ten times more frequently in the Qpar magnetosheath, primarily at $\theta_{Bn}<30^\circ$ \cite{raptis_classifying_2020, plaschke_anti-sunward_2013,vuorinen_jets_2019}. Fig.~\ref{fig:theta} shows that this strong dependence is absent at Mars. We attribute this difference to the influence of the Martian exosphere. Near perihelion the hydrogen corona is more extended, supporting the formation of PCWs upstream of the Qperp bow shock. These waves can steepen into SLAMS, driving bow shock motion and reformation even at high $\theta_{Bn}$ \cite{zhang_role_2025}. The second event shown in Fig. \ref{fig:perp} provides observational evidence consistent with this mechanism. Nevertheless, additional processes such as shock nonstationarity \cite{cheng_bow_2025} may also contribute to jet formation under Qperp conditions.

The formation of jets downstream of Qperp shocks remains poorly understood even at Earth. Jets behind the Qperp shock at Earth are found to be associated with high solar wind speeds, lower densities, lower magnetic field strengths, and higher Alfv\'en Mach numbers \cite{goncharov_evolution_2020, vuorinen_solar_2023}. These same trends are also evident at Mars (Fig.~S1) for both Qpar and Qperp jets. 

At Earth, some Qperp jets are thought to be ``encapsulated jets'', which originate at the Qpar shock but are later observed in the Qperp magnetosheath either through an IMF rotation or by traveling between the Qpar and Qperp regions \cite{raptis_classifying_2020,kajdic_causes_2021}. The smaller spatial scale of the Martian bow shock and magnetosheath may similarly allow jets formed at low $\theta_{Bn}$ to be observed at larger $\theta_{Bn}$, so some fraction of our Qperp jets may be of this origin. However, several features of our observations argue against encapsulation as the dominant mechanism: (i) the seasonal enhancement of high-$\theta_{Bn}$ jet occurrence near perihelion (Fig.~\ref{fig:theta}b) has no obvious geometric explanation under encapsulation; (ii) our analysis excluded jets with significant $\theta_{Bn}$ variation; and (iii) only $\sim5\%$ of terrestrial jets are encapsulated \cite{raptis_classifying_2020}, with no reason to expect significantly increased occurrence at Mars. The direct MAVEN detection of upstream PCWs immediately after the Qperp event (Fig.~\ref{fig:pcw}), together with simultaneous upstream SLAMS at Tianwen-1 (Fig.~\ref{fig:perp}e), supports a PCW/SLAMS-driven origin in that case.

Beyond Mars, exospheres extending into the solar wind and resulting pickup-ion populations may systematically modify bow shock processes at comets and exoplanets. Recent Rosetta observations at 67P/Churyumov--Gerasimenko have identified candidate jet events in the cometosheath \cite{goetz_density_2026}. Such comets show significant wave activity upstream of the bow shock \cite{tsurutani_comets_1991,goetz_plasma_2022}, likely leading to a similar lack of strong dependence on $\theta_{Bn}$. Modeling of Venus-like exoplanets has suggested that exoplanets can host hot coronae extending several planetary radii \cite{lee_exosphere_2021}. For weakly magnetized or unmagnetized exoplanets, the exosphere can extend into the stellar wind, and the resulting upstream waves may induce shock reformation and jets across all $\theta_{Bn}$. Upstream wave activity therefore may be a persistent source of Qperp jets at collisionless shocks in a variety of systems.

These results suggest that jet generation is influenced not only by external solar wind forcing but also by intrinsic planetary properties, exemplified by Mars' extended exosphere driving strong PCW activity upstream of the bow shock. This work highlights the importance of multi-spacecraft observations beyond Earth, including future Mars missions such as ESCAPADE, and underscores the value of comparative magnetospheric studies for understanding how similar plasma processes operate differently across planetary environments.

\matmethods{Level-2 calibrated data from the magnetometer (MAG) \cite{connerney_first_2015, connerney_maven_2015} and the Solar Wind Ion Analyzer (SWIA) \cite{halekas_solar_2015} onboard NASA's MAVEN spacecraft \cite{jakosky_mars_2015} were used. Level-2 calibrated magnetic field data from the magnetometer onboard the Chinese Tianwen-1 spacecraft (MOMAG) \cite{wang_calibration_2024,wang_mars_2023,zouflight_2023} were also used. This study used Mars–Solar–Orbital (MSO) coordinates, where $\hat{x}$ points from Mars toward the Sun, $\hat{z}$ is normal to Mars' orbital plane pointing northward, and $\hat{y} = \hat{z} \times \hat{x}$ completes the right-handed system. 

Jets were taken from the database of \cite{mohammed-amin_jets_2025}, where jets were identified using the criterion $P_{dyn} \ge 2\langle P_{dyn}\rangle_{10\,\mathrm{min}}$, where $\langle P_{dyn}\rangle_{10\,\mathrm{min}}$ denotes the 10-min background average. $P_{dyn} = m_p n v^2$, where $m_p$ and $n$ are the proton mass and number density. SWIA does not distinguish different ion species; however, Suprathermal and Thermal Ion Composition (STATIC) \cite{mcfadden_maven_2015} moments were used to assess heavy ion contributions to the total dynamic pressure, which were found to be minimal. Jets occurring within two minutes of each other were treated as a single event. A total of 1270 candidate events had simultaneous Tianwen-1 solar wind observations. The distribution of the jets in Mars' orbit is shown in Fig.~S5. The averaged magnetic field measured by Tianwen-1 during the five minutes preceding each jet was used to determine $\theta_{Bn}$, and only events satisfying $\mathrm{std}(\theta_{Bn}) < 10^\circ$ were retained. Different averaging periods were tested (5, 10, and 20 minutes) and did not significantly change the results (see Fig.~S6). The final dataset contains 733 jets, and their distribution is summarized in Table 1. 5,000 bootstrapped distributions were created from this dataset and $3\sigma$ error bars are shown for Fig.~\ref{fig:theta}.

\begin{table}\centering
\caption{Statistics of the dataset presented in Fig. 5}
    \begin{tabular}{lccc}
         & $L_s<180^\circ$ & $L_s>180^\circ$ & Total\\
         \midrule
         Qperp Jets & 372 & 106 & 478\\
         Qpar Jets & 217 & 38 & 255 \\
         All Jets & 589 & 144 & 733\\
         \bottomrule
    \end{tabular}
\end{table}

$\theta_{Bn}$ is defined as the angle between the bow shock normal $\hat{n}_{BS}$ and the IMF. Typical methods of calculating $\hat{n}_{BS}$ include minimum-variance analysis \cite{kawano_generalization_1996} or the co-planarity theorem \cite{lepping_single_1971}. These methods rely on direct traversal of the bow shock by a spacecraft, which is not available for the vast majority of events due to the positioning of MAVEN and Tianwen-1. 

Instead, the conic Trotignon et al. 2006 model \cite{trotignon_martian_2006} was used, which is based on a symmetric conic section around the $x$-axis. MAVEN's position was traced to the modeled bow shock using its solar zenith angle (SZA). This gave an $(X,Y,Z)_{MSO}$ position on the model bow shock. The bow shock normal $\hat{n}_{BS}$ was calculated as the gradient of the model at that location:

\begin{align*}
        \hat{n}_{BS} = (n_x,n_y,n_z) = (2(e^2-1)(X-X_F)-2eL,\,-2Y,\,-2Z) \numberthis
\end{align*}

\noindent Here $e=1.026$, $L=2.081$, $X_F=0.6$ are the conic section constants determined by \cite{trotignon_martian_2006}. $\theta_{Bn}$ is thus determined by: 

\begin{align*}
    \theta_{Bn} = \arccos{\frac{\hat{n}_{BS} \cdot \vec{B}_{IMF}}{|\hat{n}_{BS}||\vec{B}_{IMF}|}} \numberthis
\end{align*}

\noindent where $\vec{B}_{IMF}$ is the average IMF measured by Tianwen-1 during the five minutes preceding each jet. To assess the sensitivity of the results to the bow shock model, $\theta_{Bn}$ was recalculated using the bow shock model from Vignes et al. 2000 \cite{vignes_solar_2000}, which resulted in less than a degree of deviation. Furthermore, Fruchtman et al. \cite{fruchtman_seasonal_2023} found good agreement between the Trotignon et al. 2006 model and $\hat{n}_{BS}$ individually calculated from bow shock crossings, demonstrating their reliability.

Upstream solar wind conditions (Fig.~S1) were determined using MAVEN measurements, as Tianwen-1 lacked upstream plasma measurements. For each orbit, solar wind parameters were calculated from intervals 30 minutes before (after) magnetosheath entry (exit), provided the IMF orientation remained stable (change in IMF direction $<30^\circ$) \cite{zhang_three-dimensional_2022, zhang_energetic_2024}. Note that this dataset was assembled separately from the jet analysis where the criterion of stable IMF orientation was not applied. Kernel density estimates (KDEs) \cite{bowman_applied_1997} were used to compare the distributions of upstream solar wind parameters for all available upstream observations and for jet-producing intervals. Details of the event selection, sample sizes, bandwidth selection, and bootstrap uncertainty estimation are provided in the Supplemental Information.

}

\showmatmethods{} 

\dataavail{The research described in this manuscript utilizes publicly available data from the MAVEN mission, including data from the SWIA and MAG instruments \cite{connerney2023maven,10.17189/1414182}. Tianwen-1 MOMAG data is available at \cite{momag}. Data analysis was performed in part using the irfu-matlab software package \cite{khotyaintsev_irfu-matlab_2024}. This paper utilized the jet database from \cite{mohammed-amin_jets_2025}, which can be found at \cite{gunell_2024_14215141}.}

\acknow{This work was supported by the MAVEN Project, NASA Grants 80NSSC23K1125, 80NSSC23K0911, 80NSSC24K1843, the Alfred P. Sloan Research Fellowship, and the IBM Einstein Fellow Fund at the Institute for Advanced Study, Princeton. SR acknowledges funding from the Johns Hopkins University Applied Physics Laboratory independent R\&D fund.}

\showacknow{} 

\section*{Author Contributions}
All of the authors made notable contributions to this work. A.T., C.D., and C.Z. conceived this study and carried out the data analysis, interpretation, and manuscript preparation. S.R., H.Z., X.L., and J.G. reviewed and provided feedback on the paper. All authors contributed to the discussion and read and commented on the manuscript.

\section*{Competing Interests}
The authors declare no competing interests.


\clearpage
\onecolumn
\setcounter{figure}{0}
\renewcommand{\thefigure}{S\arabic{figure}}
\renewcommand{\theHfigure}{S\arabic{figure}}

\begin{center}
{\Large\bfseries Supplemental Information}\par\vspace{0.75em}
{\large\bfseries Widespread plasma jets in the Martian magnetosheath revealed by dual-spacecraft observations}\par\vspace{0.5em}
Abigail Tadlock, Chuanfei Dong, Chi Zhang, Savvas Raptis, Hongyang Zhou, Xinmin Li, and Jiawei Gao
\end{center}

\section*{Upstream solar wind analysis}
Upstream solar wind measurements were available for 6,338 MAVEN orbits that make up the ``All SW'' dataset in Fig.~S1. Bow shock crossings were manually identified, and upstream solar wind parameters were calculated using measurements obtained either 30 minutes after MAVEN exited the magnetosheath or 30 minutes before it entered the magnetosheath \cite{zhang_three-dimensional_2022, zhang_energetic_2024}. To ensure stable upstream conditions, only intervals with steady IMF orientations were included, defined by an angle smaller than $30^\circ$ between the inbound and outbound IMF measured by MAG. Note that this dataset was assembled separately from the jet occurrence dataset presented in Fig. 5, and therefore the steady-IMF criterion does not affect the jet occurrence statistics.

248 jets (150 with $\theta_{Bn}>45^\circ$ and 98 with $\theta_{Bn}<45^\circ$) occurred during an orbit with available upstream measurements from MAVEN. The relatively small number results from two factors: our consideration of only jets occurring after November 2021 (the arrival of Tianwen-1), and the gradual precession of MAVEN's apoapsis toward the nightside over the course of the mission, which reduced the frequency of upstream solar wind observations. As a result, the ``All SW'' dataset contains more orbits than are represented by the 248 jets. Owing to the limited number of events with upstream coverage, we did not apply the $\mathrm{std}(\theta_{Bn})$ selection criterion in this analysis; the mean standard deviation of the dataset was approximately $10^\circ$.

Kernel density estimates (KDEs) \cite{bowman_applied_1997} were used to estimate the probability density functions of jet occurrence as a function of $\theta_{Bn}$ and solar wind parameters. The KDE bandwidth was chosen to be 0.5$\sigma$ (where $\sigma$ is the standard deviation of the dataset), which provided a balance between over-smoothing and statistical noise. Uncertainties were estimated using 5,000 bootstrapped datasets; the mean distribution and its $1\sigma$ variation were then calculated.

\begin{figure}[p]
    \centering
    \includegraphics[width=\textwidth]{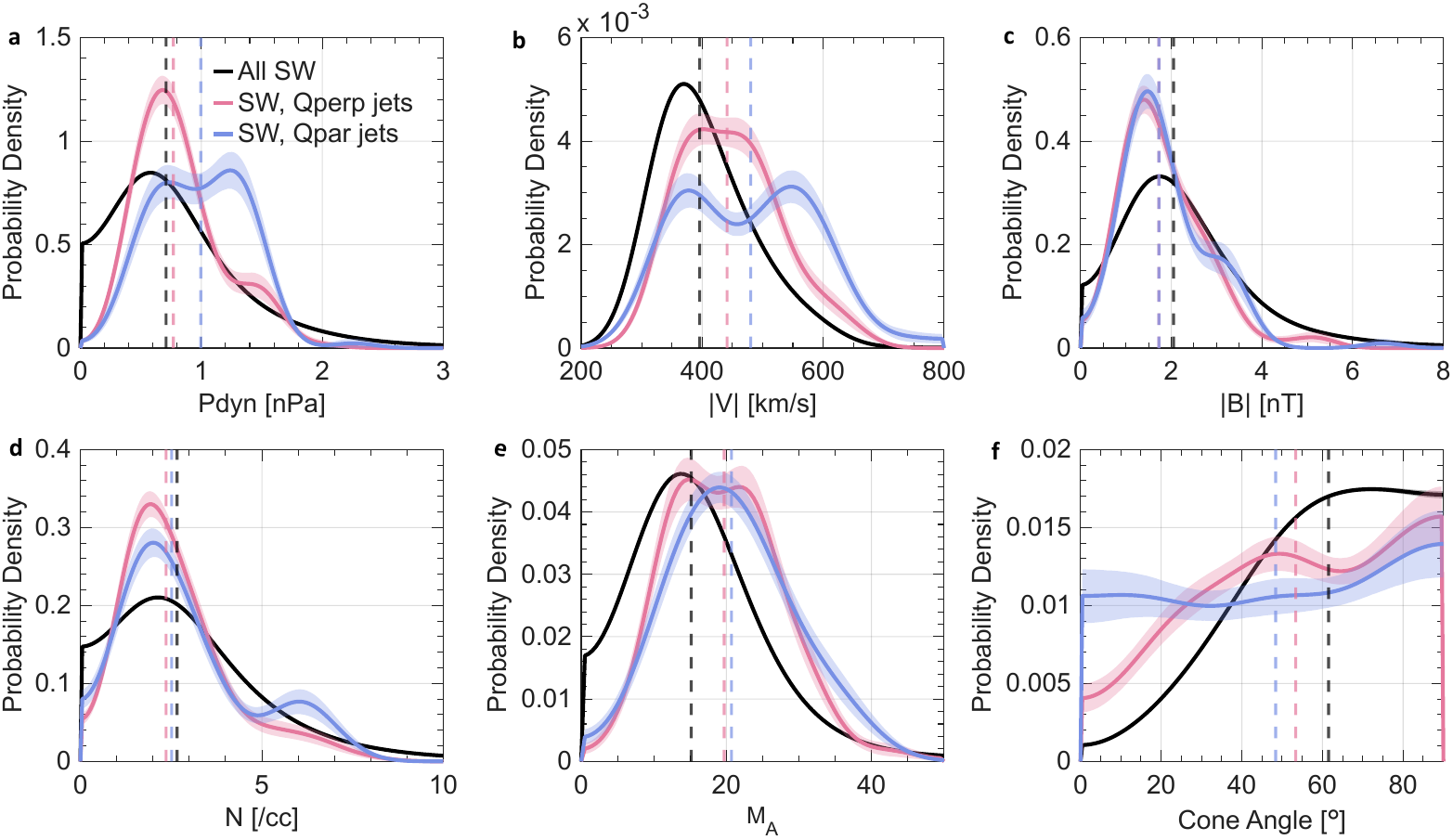}
    \caption{\textbf{Jet occurrence as a function of upstream solar wind parameters.}
    KDEs with $1\sigma$ uncertainties for upstream solar wind parameters corresponding to Qpar jets (pink), Qperp jets (blue), and the ``All SW'' dataset (black). Dashed lines indicate the median of each distribution.
    (a) Dynamic pressure $P_{dyn}$.
    (b) Solar wind speed $|\textbf{V}|$.
    (c) Magnetic field strength $|\textbf{B}|$.
    (d) Plasma density $N$.
    (e) Alfv\'en Mach number $M_A$.
    (f) IMF cone angle, $\theta = \cos^{-1}(|B_x|/|\textbf{B}|)$.
    }
    \label{fig:stat}
\end{figure}

\begin{figure}[p]
    \centering
    \includegraphics[width=\textwidth]{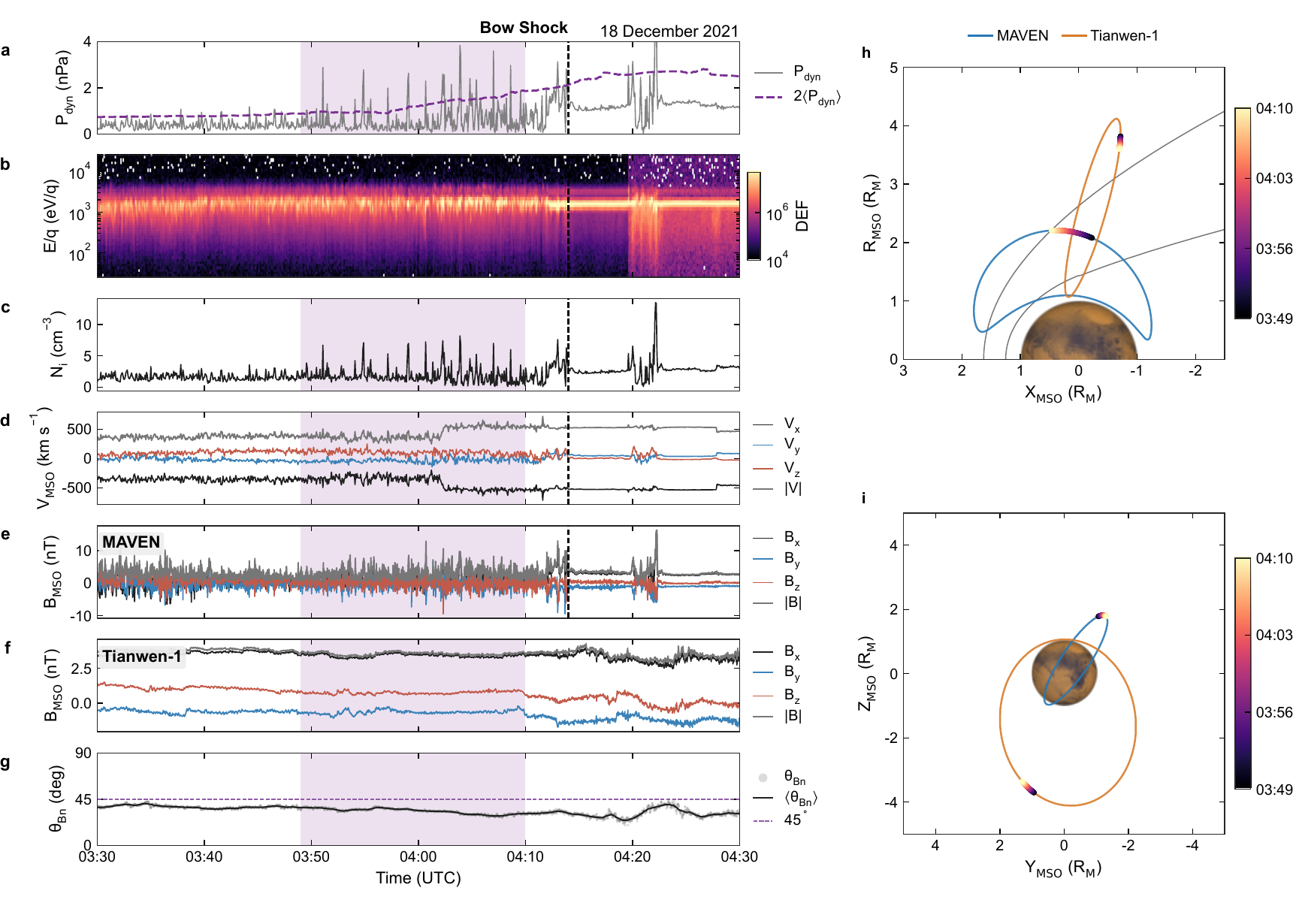}
    \caption{\textbf{Extended time series of Fig. 2.}
     (a) Dynamic pressure. The purple dashed curve represents twice the local background dynamic pressure,
    $2\left\langle P_{\mathrm{dyn}} \right\rangle$, where
    $\left\langle P_{\mathrm{dyn}} \right\rangle$ is calculated using a centered 10-min moving average.
        (b) Ion energy spectrum measured by the Solar Wind Ion Analyzer (SWIA).
    (c) Ion density measured by SWIA.
    (d) Ion velocity measured by SWIA.
    (e) Magnetic field measured by MAVEN.
    (f) Magnetic field measured by Tianwen-1.
    (g) $\theta_{Bn}$ calculated using the IMF measured by Tianwen-1 and the position of MAVEN; the black curve shows a 1-min moving average.
    (h) Trajectories and positions of MAVEN (blue) and Tianwen-1 (orange) in the cylindrical X--R plane.
    (i) Positions of MAVEN and Tianwen-1 in the Y--Z MSO plane.
    The MAVEN bow shock crossing is indicated by the vertical dashed black line in panels (a)--(e). Tianwen-1 remains in the solar wind throughout the period shown. The time interval presented in Fig. 2 is highlighted by the shaded region.}
    \label{fig:s1}
\end{figure}

\begin{figure}[p]
    \centering
    \includegraphics[width=\textwidth]{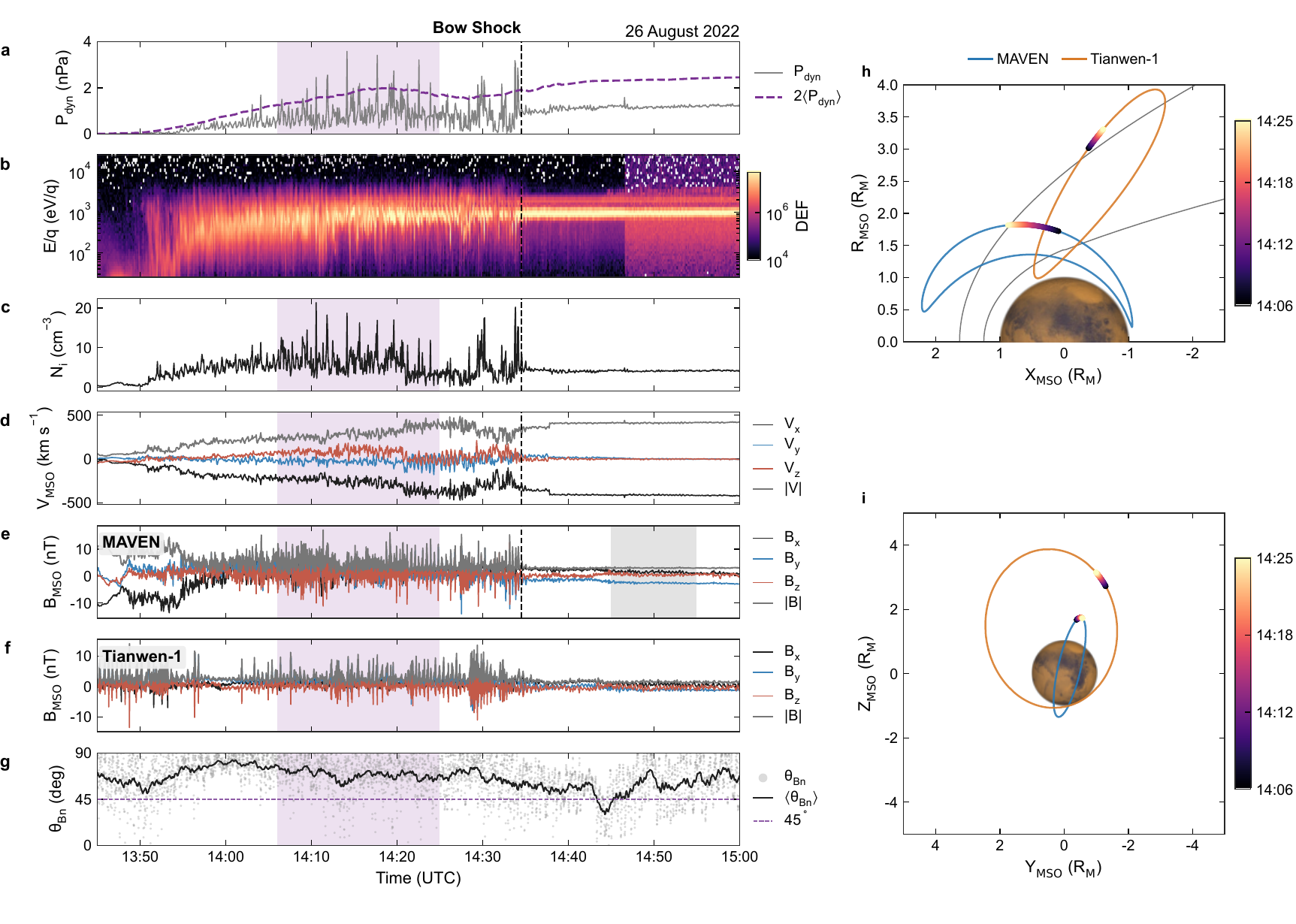}
    \caption{\textbf{Extended version of Fig. 3.}
    Extended time series in the same format as Fig. S2. The MAVEN bow shock crossing is indicated by the vertical dashed black line in panels (a)--(e). The time interval shown in Fig. 3 is highlighted by the left shaded region. The interval analyzed in Fig. 4 of the main text is highlighted by the right shaded region in panel (e).
    }
    \label{fig:s3}
\end{figure}

\begin{figure}[p]
    \centering
    \includegraphics[width=0.7\linewidth]{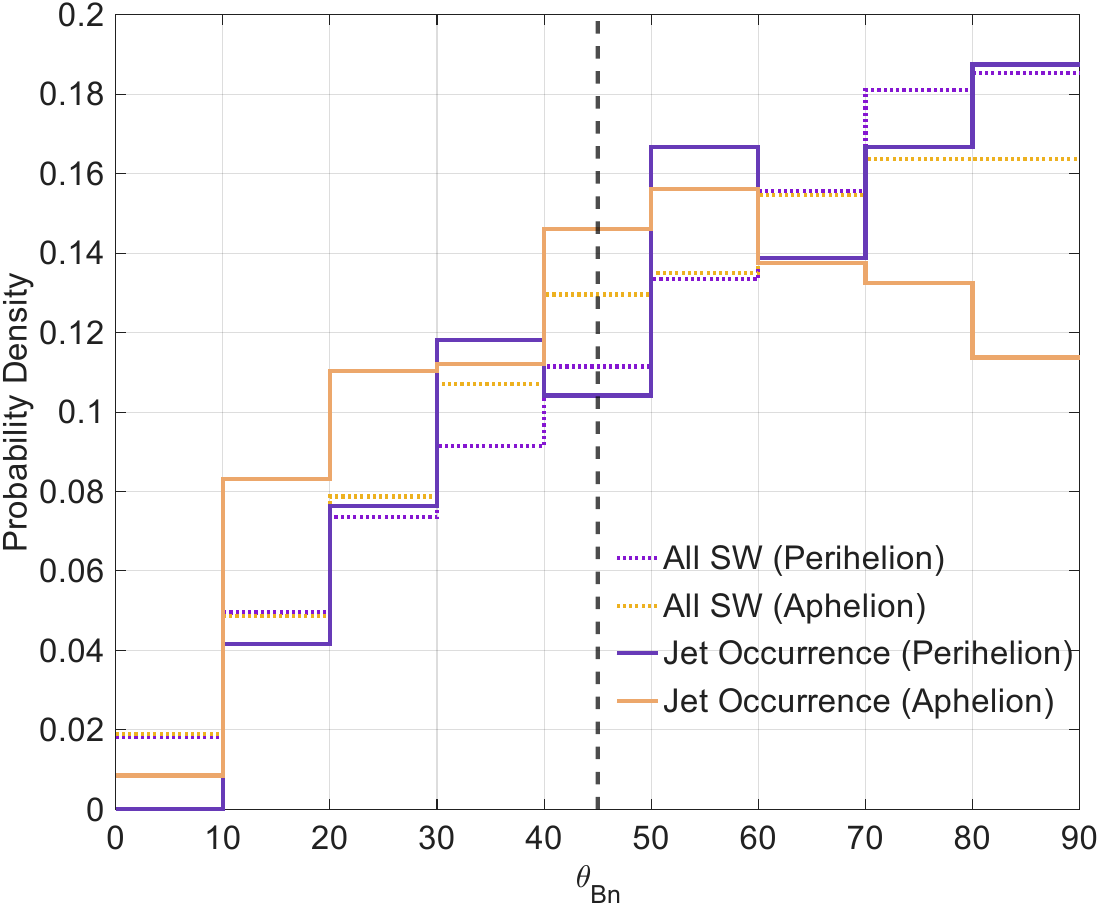}
    \caption{\textbf{Breakdown of magnetosheath $\theta_{Bn}$ observations.} Same as Figure 5a, but with the ``All SW'' distribution divided by solar longitude in the same manner as the jet observations. The jet observations were normalized by these subdivided background distributions, not the black combined distribution shown in Fig. 5a.}
    \label{fig:s4}
\end{figure}

\begin{figure}[p]
    \centering
    \includegraphics[width=0.5\linewidth]{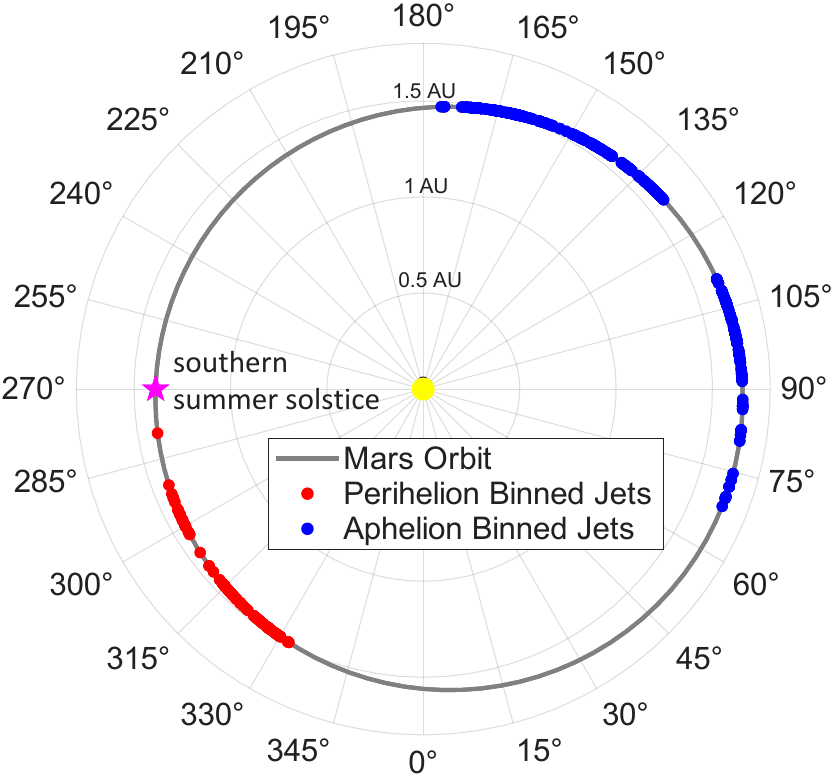}
    \caption{\textbf{Orbital distribution of observations.} The distribution of the 733 jets included in main text Fig. 5 shown along Mars' orbit as a function of solar longitude ($L_s$); the radial axis is the heliocentric distance in AU. Jets are shown separately for the near-aphelion (blue) and near-perihelion (red) bins. The gaps in coverage are due to periods without available Tianwen-1 data. The southern summer solstice ($L_s=270^\circ$), when exospheric column density peaks \cite{clarke_martian_2024, halekas_seasonal_2017}, is marked with a purple star.}
    \label{fig:s5}
\end{figure}

\begin{figure}[p]
    \centering
    \includegraphics[width=0.9\linewidth]{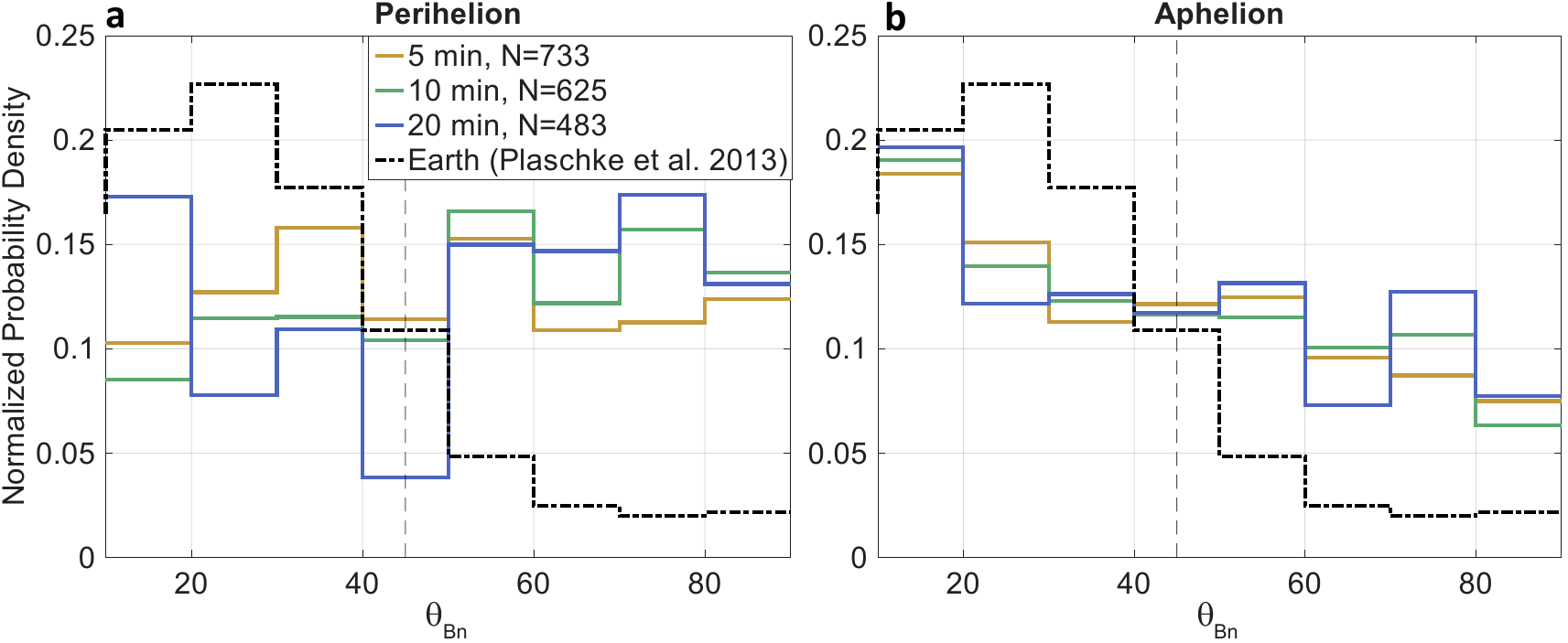}
    \caption{\textbf{Comparison of different IMF averaging windows.} Jet occurrence as a function of $\theta_{Bn}$ in the same manner as main text Fig. 5b, with $\theta_{Bn}$ calculated using different averaging window lengths. The panels are (a) near-perihelion and (b) near-aphelion. The number of events $N$ is denoted in the legend. The results are consistent with the 5-min averaging window used in Fig. 5.}
    \label{fig:s6}
\end{figure}

\clearpage


\begin{thebibliography}{10}

\bibitem{ma_comprehensive_2024}
J Ma, et~al., The {Comprehensive} {Response} of the {Magnetopause} to the
  {Impact} of an {Isolated} {Magnetosheath} {High}-{Speed} {Jet}.
\newblock {\em\protect\JournalTitle{Geophysical Research Letters}} \textbf{51},
  e2024GL111132 (2024).

\bibitem{amata_high_2011}
E Amata, et~al., High kinetic energy density jets in the {Earth}’s
  magnetosheath: {A} case study.
\newblock {\em\protect\JournalTitle{Planetary and Space Science}} \textbf{59},
  482--494 (2011).

\bibitem{archer_direct_2019}
MO Archer, H Hietala, MD Hartinger, F Plaschke, V Angelopoulos, Direct
  observations of a surface eigenmode of the dayside magnetopause.
\newblock {\em\protect\JournalTitle{Nature Communications}} \textbf{10}, 615
  (2019).

\bibitem{archer_magnetospheric_2013}
MO Archer, TS Horbury, JP Eastwood, JM Weygand, TK Yeoman, Magnetospheric
  response to magnetosheath pressure pulses: {A} low-pass filter effect.
\newblock {\em\protect\JournalTitle{Journal of Geophysical Research: Space
  Physics}} \textbf{118}, 5454--5466 (2013).

\bibitem{wang_impacts_2018}
B Wang, et~al., Impacts of {Magnetosheath} {High}-{Speed} {Jets} on the
  {Magnetosphere} and {Ionosphere} {Measured} by {Optical} {Imaging} and
  {Satellite} {Observations}.
\newblock {\em\protect\JournalTitle{Journal of Geophysical Research: Space
  Physics}} \textbf{123}, 4879--4894 (2018).

\bibitem{plaschke_jets_2018}
F Plaschke, et~al., Jets {Downstream} of {Collisionless} {Shocks}.
\newblock {\em\protect\JournalTitle{Space Science Reviews}} \textbf{214}, 81
  (2018).

\bibitem{kramer_jets_2024}
E Krämer, et~al., Jets {Downstream} of {Collisionless} {Shocks}: {Recent}
  {Discoveries} and {Challenges}.
\newblock {\em\protect\JournalTitle{Space Science Reviews}} \textbf{221}, 4
  (2024).

\bibitem{plaschke_geoeffective_2016}
F Plaschke, H Hietala, V Angelopoulos, R Nakamura, Geoeffective jets impacting
  the magnetopause are very common.
\newblock {\em\protect\JournalTitle{Journal of Geophysical Research: Space
  Physics}} \textbf{121}, 3240--3253 (2016).

\bibitem{hietala_supermagnetosonic_2009}
H Hietala, et~al., Supermagnetosonic {Jets} behind a {Collisionless}
  {Quasiparallel} {Shock}.
\newblock {\em\protect\JournalTitle{Physical Review Letters}} \textbf{103},
  245001 (2009).

\bibitem{plaschke_anti-sunward_2013}
F Plaschke, H Hietala, V Angelopoulos, Anti-sunward high-speed jets in the
  subsolar magnetosheath.
\newblock {\em\protect\JournalTitle{Annales Geophysicae}} \textbf{31},
  1877--1889 (2013).

\bibitem{vuorinen_jets_2019}
L Vuorinen, H Hietala, F Plaschke, Jets in the magnetosheath: {IMF} control of
  where they occur.
\newblock {\em\protect\JournalTitle{Annales Geophysicae}} \textbf{37}, 689--697
  (2019).

\bibitem{raptis_classifying_2020}
S Raptis, T Karlsson, F Plaschke, A Kullen, PA Lindqvist, Classifying
  {Magnetosheath} {Jets} {Using} {MMS}: {Statistical} {Properties}.
\newblock {\em\protect\JournalTitle{Journal of Geophysical Research: Space
  Physics}} \textbf{125}, e2019JA027754 (2020).

\bibitem{archer_magnetosheath_2012}
MO Archer, TS Horbury, JP Eastwood, Magnetosheath pressure pulses: {Generation}
  downstream of the bow shock from solar wind discontinuities.
\newblock {\em\protect\JournalTitle{Journal of Geophysical Research: Space
  Physics}} \textbf{117}, A05228 (2012).

\bibitem{suni_magnetosheath_2025}
J Suni, et~al., Magnetosheath {Jets} {Associated} {With} a {Solar} {Wind}
  {Rotational} {Discontinuity} in a {Hybrid}-{Vlasov} {Simulation}.
\newblock {\em\protect\JournalTitle{Journal of Geophysical Research: Space
  Physics}} \textbf{130}, e2025JA033995 (2025).

\bibitem{hietala_generation_2013}
H Hietala, F Plaschke, On the generation of magnetosheath high-speed jets by
  bow shock ripples.
\newblock {\em\protect\JournalTitle{Journal of Geophysical Research: Space
  Physics}} \textbf{118}, 7237--7245 (2013).

\bibitem{raptis_downstream_2022}
S Raptis, et~al., Downstream high-speed plasma jet generation as a direct
  consequence of shock reformation.
\newblock {\em\protect\JournalTitle{Nature Communications}} \textbf{13}, 598
  (2022).

\bibitem{gunell_magnetosheath_2023}
H Gunell, M Hamrin, S Nesbit-Östman, E Krämer, H Nilsson, Magnetosheath jets
  at {Mars}.
\newblock {\em\protect\JournalTitle{Science Advances}} \textbf{9}, eadg5703
  (2023).

\bibitem{zhou_magnetosheath_2024}
Y Zhou, et~al., Magnetosheath jets at {Jupiter} and across the solar system.
\newblock {\em\protect\JournalTitle{Nature Communications}} \textbf{15}, 4
  (2024).

\bibitem{zhang_three-dimensional_2022}
C Zhang, et~al., Three-{Dimensional} {Configuration} of {Induced} {Magnetic}
  {Fields} {Around} {Mars}.
\newblock {\em\protect\JournalTitle{Journal of Geophysical Research: Planets}}
  \textbf{127}, e2022JE007334 (2022).

\bibitem{gao_two_2024}
J Gao, et~al., Two distinct current systems in the ionosphere of {Mars}.
\newblock {\em\protect\JournalTitle{Nature Communications}} \textbf{15}, 9704
  (2024).

\bibitem{nagy_plasma_2004}
A Nagy, et~al., The plasma {Environment} of {Mars}.
\newblock {\em\protect\JournalTitle{Space Science Reviews}} \textbf{111},
  33--114 (2004).

\bibitem{dong_solar_2015}
C Dong, et~al., Solar wind interaction with the {Martian} upper atmosphere:
  {Crustal} field orientation, solar cycle, and seasonal variations.
\newblock {\em\protect\JournalTitle{Journal of Geophysical Research: Space
  Physics}} \textbf{120}, 7857--7872 (2015).

\bibitem{mohammed-amin_jets_2025}
T Mohammed-Amin, E Krämer, S Nesbit-Östman, H Gunell, CS Wedlund, Jets
  downstream of the {Martian} bow shock - {Occurrence} in the 2014–2024
  period.
\newblock {\em\protect\JournalTitle{Astronomy \& Astrophysics}} \textbf{696},
  A75 (2025).

\bibitem{wei_proton_2006}
HY Wei, CT Russell, Proton cyclotron waves at {Mars}: {Exosphere} structure and
  evidence for a fast neutral disk.
\newblock {\em\protect\JournalTitle{Geophysical Research Letters}} \textbf{33},
  L23103 (2006).

\bibitem{zhang_role_2025}
C Zhang, et~al., Role of {ULF} {Waves} in {Reforming} the {Martian} {Bow}
  {Shock}.
\newblock {\em\protect\JournalTitle{AGU Advances}} \textbf{6}, e2025AV001654
  (2025).

\bibitem{mazelle_bow_2004}
C Mazelle, et~al., Bow {Shock} and {Upstream} {Phenomena} at {Mars}.
\newblock {\em\protect\JournalTitle{Space Science Reviews}} \textbf{111},
  115--181 (2004).

\bibitem{delva_upstream_2011}
M Delva, C Mazelle, C Bertucci, Upstream {Ion} {Cyclotron} {Waves} at {Venus}
  and {Mars}.
\newblock {\em\protect\JournalTitle{Space Science Reviews}} \textbf{162}, 5--24
  (2011).

\bibitem{romanelli_proton_2016}
N Romanelli, et~al., Proton cyclotron waves occurrence rate upstream from
  {Mars} observed by {MAVEN}: {Associated} variability of the {Martian} upper
  atmosphere.
\newblock {\em\protect\JournalTitle{Journal of Geophysical Research: Space
  Physics}} \textbf{121}, 11,113--11,128 (2016).

\bibitem{jakosky_mars_2015}
BM Jakosky, et~al., The {Mars} {Atmosphere} and {Volatile} {Evolution}
  ({MAVEN}) {Mission}.
\newblock {\em\protect\JournalTitle{Space Science Reviews}} \textbf{195}, 3--48
  (2015).

\bibitem{zou_scientific_2021}
Y Zou, et~al., Scientific objectives and payloads of {Tianwen}-1, {China}’s
  first {Mars} exploration mission.
\newblock {\em\protect\JournalTitle{Advances in Space Research}} \textbf{67},
  812--823 (2021).

\bibitem{romeo_variability_2021}
OM Romeo, et~al., Variability of {Upstream} {Proton} {Cyclotron} {Wave}
  {Properties} and {Occurrence} at {Mars} {Observed} by {MAVEN}.
\newblock {\em\protect\JournalTitle{Journal of Geophysical Research: Space
  Physics}} \textbf{126}, e2020JA028616 (2021).

\bibitem{halekas_seasonal_2017}
JS Halekas, Seasonal variability of the hydrogen exosphere of {Mars}.
\newblock {\em\protect\JournalTitle{Journal of Geophysical Research: Planets}}
  \textbf{122}, 901--911 (2017).

\bibitem{clarke_martian_2024}
JT Clarke, et~al., Martian atmospheric hydrogen and deuterium: Seasonal changes
  and paradigm for escape to space.
\newblock {\em\protect\JournalTitle{Science Advances}} \textbf{10}, eadm7499
  (2024).

\bibitem{trotignon_martian_2006}
JG Trotignon, C Mazelle, C Bertucci, MH Acuña, Martian shock and magnetic
  pile-up boundary positions and shapes determined from the {Phobos} 2 and
  {Mars} {Global} {Surveyor} data sets.
\newblock {\em\protect\JournalTitle{Planetary and Space Science}} \textbf{54},
  357--369 (2006).

\bibitem{liu_statistical_2021}
D Liu, et~al., Statistical {Properties} of {Solar} {Wind} {Upstream} of {Mars}:
  {MAVEN} {Observations}.
\newblock {\em\protect\JournalTitle{The Astrophysical Journal}} \textbf{911},
  113 (2021).

\bibitem{yamauchi_seasonal_2015}
M Yamauchi, et~al., Seasonal variation of {Martian} pick-up ions: {Evidence} of
  breathing exosphere.
\newblock {\em\protect\JournalTitle{Planetary and Space Science}} \textbf{119},
  54--61 (2015).

\bibitem{karlsson_origin_2015}
T Karlsson, et~al., On the origin of magnetosheath plasmoids and their relation
  to magnetosheath jets.
\newblock {\em\protect\JournalTitle{Journal of Geophysical Research: Space
  Physics}} \textbf{120}, 7390--7403 (2015).

\bibitem{lamoury_solar_2021}
AT LaMoury, H Hietala, F Plaschke, L Vuorinen, JP Eastwood, Solar {Wind}
  {Control} of {Magnetosheath} {Jet} {Formation} and {Propagation} to the
  {Magnetopause}.
\newblock {\em\protect\JournalTitle{Journal of Geophysical Research: Space
  Physics}} \textbf{126}, e2021JA029592 (2021).

\bibitem{vuorinen_solar_2023}
L Vuorinen, H Hietala, AT LaMoury, F Plaschke, Solar {Wind} {Parameters}
  {Influencing} {Magnetosheath} {Jet} {Formation}: {Low} and {High} {IMF}
  {Cone} {Angle} {Regimes}.
\newblock {\em\protect\JournalTitle{Journal of Geophysical Research: Space
  Physics}} \textbf{128}, e2023JA031494 (2023).

\bibitem{cheng_bow_2025}
L Cheng, et~al., Bow shock oscillations of {Mars} under weakly disturbed solar
  wind conditions.
\newblock {\em\protect\JournalTitle{Nature Communications}} \textbf{16}, 9649
  (2025).

\bibitem{goncharov_evolution_2020}
O Goncharov, H Gunell, M Hamrin, S Chong, Evolution of {High}-{Speed} {Jets}
  and {Plasmoids} {Downstream} of the {Quasi}-{Perpendicular} {Bow} {Shock}.
\newblock {\em\protect\JournalTitle{Journal of Geophysical Research: Space
  Physics}} \textbf{125}, e2019JA027667 (2020).

\bibitem{kajdic_causes_2021}
P Kajdič, S Raptis, X Blanco-Cano, T Karlsson, Causes of {Jets} in the
  {Quasi}-{Perpendicular} {Magnetosheath}.
\newblock {\em\protect\JournalTitle{Geophysical Research Letters}} \textbf{48},
  e2021GL093173 (2021).

\bibitem{goetz_density_2026}
C Goetz, et~al., Density {Increases} in the {Cometary} {Plasma} {Environment}
  and {Their} {Relation} to {Magnetosheath} {Jets}.
\newblock {\em\protect\JournalTitle{The Astrophysical Journal}} \textbf{1006},
  62 (2026).

\bibitem{tsurutani_comets_1991}
BT Tsurutani, Comets: a laboratory for plasma waves and instabilities, in
  {\em Cometary Plasma Processes}, ed. A Johnstone, Geophysical Monograph
  Series, Vol. 61.
\newblock (American Geophysical Union, Washington, DC, 1991), pp. 189--209.

\bibitem{goetz_plasma_2022}
C Goetz, et~al., The {Plasma} {Environment} of {Comet}
  {67P}/{Churyumov}-{Gerasimenko}.
\newblock {\em\protect\JournalTitle{Space Science Reviews}} \textbf{218}, 65
  (2022).

\bibitem{lee_exosphere_2021}
Y Lee, C Dong, V Tenishev, Exosphere {Modeling} of {Proxima} b: {A} {Case}
  {Study} of {Photochemical} {Escape} with a {Venus}-like {Atmosphere}.
\newblock {\em\protect\JournalTitle{The Astrophysical Journal}} \textbf{923},
  190 (2021).

\bibitem{connerney_first_2015}
JEP Connerney, et~al., First results of the {MAVEN} magnetic field
  investigation.
\newblock {\em\protect\JournalTitle{Geophysical Research Letters}} \textbf{42},
  8819--8827 (2015).

\bibitem{connerney_maven_2015}
JEP Connerney, et~al., The {MAVEN} {Magnetic} {Field} {Investigation}.
\newblock {\em\protect\JournalTitle{Space Science Reviews}} \textbf{195},
  257--291 (2015).

\bibitem{halekas_solar_2015}
JS Halekas, et~al., The {Solar} {Wind} {Ion} {Analyzer} for {MAVEN}.
\newblock {\em\protect\JournalTitle{Space Science Reviews}} \textbf{195},
  125--151 (2015).

\bibitem{wang_calibration_2024}
G Wang, et~al., Calibration of the {Zero} {Offset} of the {Fluxgate}
  {Magnetometer} on {Board} the {Tianwen}-1 {Orbiter} in the {Martian}
  {Magnetosheath}.
\newblock {\em\protect\JournalTitle{Journal of Geophysical Research: Space
  Physics}} \textbf{129}, e2023JA031757 (2024).

\bibitem{wang_mars_2023}
Y Wang, et~al., The {Mars} orbiter magnetometer of {Tianwen}-1: in-flight
  performance and first science results.
\newblock {\em\protect\JournalTitle{Earth and Planetary Physics}} \textbf{7},
  216--228 (2023).

\bibitem{zouflight_2023}
Z Zou, et~al., In-flight calibration of the magnetometer on the {Mars} orbiter
  of {Tianwen}-1.
\newblock {\em\protect\JournalTitle{Science China Technological Sciences}}
  \textbf{66}, 2396--2405 (2023).

\bibitem{mcfadden_maven_2015}
JP McFadden, et~al., {MAVEN} {SupraThermal} and {Thermal} {Ion} {Composition}
  ({STATIC}) {Instrument}.
\newblock {\em\protect\JournalTitle{Space Science Reviews}} \textbf{195},
  199--256 (2015).

\bibitem{kawano_generalization_1996}
H Kawano, T Higuchi, A generalization of the minimum variance analysis method.
\newblock {\em\protect\JournalTitle{Annales Geophysicae}} \textbf{14},
  1019--1024 (1996).

\bibitem{lepping_single_1971}
RP Lepping, PD Argentiero, Single spacecraft method of estimating shock
  normals.
\newblock {\em\protect\JournalTitle{Journal of Geophysical Research}} \textbf{76},
  4349--4359 (1971).

\bibitem{vignes_solar_2000}
D Vignes, et~al., The solar wind interaction with {Mars}: {Locations} and
  shapes of the bow shock and the magnetic pile-up boundary from the
  observations of the {MAG}/{ER} {Experiment} onboard {Mars} {Global}
  {Surveyor}.
\newblock {\em\protect\JournalTitle{Geophysical Research Letters}} \textbf{27},
  49--52 (2000).

\bibitem{fruchtman_seasonal_2023}
J Fruchtman, J Halekas, J Gruesbeck, D Mitchell, C Mazelle, Seasonal and {Mach}
  {Number} {Variation} of the {Martian} {Bow} {Shock} {Structure}.
\newblock {\em\protect\JournalTitle{Journal of Geophysical Research: Space
  Physics}} \textbf{128}, e2023JA031759 (2023).

\bibitem{zhang_energetic_2024}
C Zhang, et~al., The {Energetic} {Oxygen} {Ion} {Beams} in the {Martian}
  {Magnetotail} {Current} {Sheets}: {Hints} {From} the {Comparisons} {Between}
  {Two} {Types} of {Current} {Sheets}.
\newblock {\em\protect\JournalTitle{Geophysical Research Letters}} \textbf{51},
  e2023GL107190 (2024).

\bibitem{bowman_applied_1997}
AW Bowman, A Azzalini, {\em Applied {Smoothing} {Techniques} for {Data}
  {Analysis}: {The} {Kernel} {Approach} with {S}-{Plus} {Illustrations}},
  Oxford {Statistical} {Science} {Series}.
\newblock (Oxford University Press, Oxford and New York, 1997).

\bibitem{connerney2023maven}
JEP Connerney, {MAVEN Magnetometer (MAG) Calibrated Data Bundle} [{Dataset}].
\newblock NASA Planetary Data System (2023).
\newblock \url{https://doi.org/10.17189/1414178}.

\bibitem{10.17189/1414182}
JS Halekas, {MAVEN SWIA Calibrated Data Bundle} [{Dataset}].
\newblock NASA Planetary Data System (2024).
\newblock \url{https://doi.org/10.17189/1414182}.

\bibitem{momag}
DREAMS, Archive of calibrated {Tianwen-1/MOMAG} data [{Dataset}] (2026).
\newblock \url{https://space.ustc.edu.cn/dreams/tw1_momag/?magdata=cal}.

\bibitem{khotyaintsev_irfu-matlab_2024}
Y Khotyaintsev, et~al., {IRFU-Matlab} [{Software}].
\newblock Zenodo (2024).
\newblock \url{https://doi.org/10.5281/zenodo.14525047}.

\bibitem{gunell_2024_14215141}
H Gunell, E Krämer, S Nesbit-Östman, C Simon~Wedlund, T Mohammed-Amin, Jets
  downstream of the {Martian} bow shock: Occurrence during the 2014--2024 period
  [{Dataset}].
\newblock Zenodo (2024).
\newblock \url{https://doi.org/10.5281/zenodo.14215141}.

\end{thebibliography}

\end{document}